\documentstyle[aps,multicol,epsf,prl]{revtex}
\begin{document}
\title{Numerical Study of Order in a Gauge Glass Model}
\author{J.M.Kosterlitz and N.Akino}
\address{Department of Physics, Brown University, Providence, RI 02912, USA}
\date{\today}
\maketitle
\draft
\begin{abstract}
The $XY$ model with quenched random phase shifts is studied by a $T=0$ finite
size defect energy scaling method in $2d$ and $3d$. The defect energy is
defined by a change in the boundary conditions from those compatible with
the true ground state configuration for a given realization of disorder. A
numerical technique, which is exact in principle, is used to evaluate this
energy and to estimate the stiffness exponent $\theta$. This method gives
$\theta =-0.36\pm 0.013$ in $2d$ and $\theta =+0.31\pm 0.015$ in $3d$, which are
considerably larger than previous estimates, strongly suggesting that the
lower critical dimension is less than three. Some arguments in favor of these
new estimates are given. 
\end{abstract}

\begin{multicols}{2}
  The $XY$ model with random quenched disorder, also known as a gauge glass, as 
a model for a superconducting glass phase has been intensively investigated over 
the last decade. Since the mean field theory flux lattice moves in response 
to a transport current, the existence of disorder, which, however,
destroys the flux lattice structure, becomes essential 
in order to have a superconducting phase in a high $T_{c}$ superconductor. 
Since the pinning of flux lines by random disorder can prevent dissipation,
a supercurrent can exist in the glass 
phase \cite{vortex-glass1,vortex-glass2}. 
\newline\indent
 From numerical\cite{gg-simulation1,gg-simulation2,gg-simulation3,gg-simulation4,gg-simulation5,gg-simulation6}
  and experimental \cite{experiment1}
  studies, it is believed that the gauge glass has no ordered phase at any finite
  temperature in two dimensions. 
 In three dimensions, numerical studies indicate
that the lower critical dimension seems to be close to three.
However the situation is less conclusive, since the simulations are 
limited to small system sizes.
Experimentally there is also some 
evidence which supports a finite temperature phase transition to 
a superconducting glass phase \cite{experiment2,experiment3}.
\newline\indent 
  The Hamiltonian of the gauge glass model is given by
 \begin{equation}
 \label{eq:h}
H=\sum_{<ij>}V(\theta_{i}-\theta_{j}-A_{ij})
\end{equation}
where $V(\phi)$ is a $2\pi$ periodic function of $\phi$ with a maximum at $\phi=\pi$,
usually taken to be $V(\phi_{ij})=-J_{ij}cos(\phi_{ij})$.
$\theta_{i}$ is the phase of the superconducting order parameter 
at site $i$ of a square lattice in $2d$ and a simple cubic lattice in $3d$. 
The sum is taken over all nearest neighbor sites. The random bond variables, 
$A_{ij}$, which are responsible for the random 
frustration, are uniformly distributed in $(-\pi,\pi]$. The coupling 
constants, $J_{ij}$, are taken as constant, $J_{ij}=J$. In this paper, we use
a domain wall renormalization group (DWRG)\cite{DWRG1,DWRG2}, or defect energy scaling, method
to investigate the possibility of an ordered phase at small but finite temperature
$T$. The idea is to find the ground state (GS) energy $E_{0}(L)$ of a system of linear size
$L$, then to change the boundary conditions (BC) in one direction to introduce a defect or domain
wall into the system and finally to find the energy $E_{D}(L)$ of this. The energy
difference $\Delta E(L)\equiv E_{D}(L)-E_{0}(L)$ is then the energy of a domain wall 
(defect) of length $L$. This is assumed to scale as
 \begin{equation}
\label{eq:deltae}
<\Delta E(L)> \sim L^{\theta}
\end{equation}
where $<\cdots>$ means an average over disorder and $\theta$ is the stiffness exponent. The
sign of $\theta$ determines if a low $T$ ordered phase exists or not. If $\theta <0$,
in the thermodynamic limit the defect energy vanishes and at any $T>0$ will 
proliferate and destroy the order. On the other hand, if $\theta >0$, the defect
energy diverges as $L\rightarrow\infty$ and the system will be ordered at sufficiently
low $T$ and one expects a phase transition at some $T_{c}>0$ between two distinct
phases. The standard method to implement these ideas is to simply find the energies
of the system subject to periodic and antiperiodic BC, despite the fact that neither
BC is compatible with the true GS and that the energy difference is not a defect
energy but is the energy difference between two randomly chosen defects. Nevertheless,
$<\Delta E(L)> \equiv  <|E_{ap}(L) - E_{p}(L)|>$
is assumed to scale as $L^{\theta}$ and the best fit to this is taken as yielding the
exponent $\theta$.
\newline\indent
In this paper, we take the point of view that, since randomness restricts the
accessible system sizes $L$ to rather small values, for the DWRG to be successful it 
is important to minimize uncontrolled and poorly understood effects which may introduce
large corrections to scaling. Also when fitting a continuous function to less than
ten data points there is scope for obtaining erroneous results. To attempt to minimize
such effects, we first transform the problem described by eq.(\ref{eq:h}) to a
Coulomb gas representation which eliminates the spin wave contribution to the energies
and also enables us to estimate numerically the true GS energy by optimizing the BC. In
the phase representation this would involve applying phase differences across all
the bonds beween opposite faces of the system on a hypertorus in $d$ dimensions 
and minimizing the energy with respect to $L^{d}$ phases and $dL^{d-1}$
phase differences on the bonds connecting opposite faces. In the Coulomb gas
representation, one needs only $d$ extra parameters, one for each direction, and
minimize with respect to $(d+L^{d})$ parameters. To introduce a domain wall, we have
to change only one of the $d$ parameters from its value in the GS and minimize the
energy with respect to $L^{d}$ bulk variables keeping the BC fixed
to find the defect energy $E_{D}$. We then obtain the {\it{true}}
domain wall energy as $\Delta E(L)\equiv E_{D}(L)-E_{0}(L) \ge 0$
for a particular realization of disorder. This procedure is repeated for several
realizations of disorder and system sizes $L$ to
obtain $<\Delta E(L)>$ which is fitted to eq.(\ref{eq:deltae}).
\newline\indent
Some support for this point of view comes from a conjecture made by 
Ney-Nifle and Hilhorst\cite{nh} in the context of the $2d$ $XY$ spin glass, which is a
special case of the model of eq.(\ref{eq:h}) when the bond variables $A_{ij}=0,\pi$ each
with probability $1/2$, based on some analytic work by Ney-Nifle, Hilhorst and Moore\cite 
{nhm} on the $XY$ spin glass on a ladder. The latter authors solve the problem analytically
and, by imposing reflective BC which forces a chiral domain wall into the system, find that
both the spin and chiral domain wall energies scale with the {\it{same}} exponent 
$\theta_{s}=\theta_{c}=-1.7972\cdots$. The same conclusion is reached in ref.\cite{nh2} but with 
a slightly different value of $\theta$. In ref.\cite{nh}, supported by heuristic but plausible
arguments, it is conjectured that, for any spin glass system below the lower critical
dimension $d_{l}$, $\theta_{s}=\theta_{c}<0$. This result, although lacking rigorous proof,
provides an important bench test for a numerical simulation which seems to have been either
ignored or overlooked in some recent studies eg.\cite{xysg1} where it is found that $\theta_{s}
<\theta_{c}$ in $d=2<d_{l}$. On the other hand, we have   
recently studied the $2d$ $XY$ spin glass using the philosophy outlined above and find that
$\theta_{s}=\theta_{c}\approx -0.35$ to the accuracy of our simulations\cite{ka2}. 
This supports the point of view taken in this paper that defect energies should be
measured by changing the BC in a controlled way from those BC consistent with the true
ground state of a system with a particular realization of randomness. If this is not done,
as in the standard method of comparing energies with periodic and antiperiodic BC, one is
not measuring a specific defect energy and the $<\Delta E(L)>$ obtained this way either does
not scale as $L^{\theta}$ or $\theta$ is the stiffness exponent of a quantity
whose meaning is unclear. In any event, if the latter scenario holds, there is no reason 
to expect any relation between $\theta_{s}$ and $\theta_{c}$. Since the $XY$ spin glass is a 
special case of the gauge glass, we expect that these considerations also hold for the latter 
system. 
\newline\indent    
The same cannot be said for the standard RT method where there
are no checks, just a hope, that true minima have been achieved.To eliminate the effect of smooth variations in the phases $\theta_{i}$, we transform
the Hamiltonian from the phase representation of eq.(\ref{eq:h}) to a Coulomb gas
representation which is more convenient for numerical work. The function $V(\phi)$ is
taken to be a piecewise parabolic potential which is equivalent to a Villain\cite{villain}
potential at $T=0$
\begin{eqnarray}
H &=& \frac{J}{2} \sum_{<ij>}(\theta_{i}-\theta_{j}-A_{ij}-2 \pi n_{ij} )^{2}
\cr 
&\equiv &\frac{J}{2} \sum_{<ij>}(\phi_{ij}-A_{ij})^{2} 
\end{eqnarray}
By a duality transformation \cite{nh,duality1,duality2}, 
in $2d$ the Coulomb gas Hamiltonian becomes
\begin{eqnarray}
\label{eq:h2d}
H=&2& \pi ^{2} J \sum_{{\bf r},{\bf r}'} (q_{\bf r} - f_{\bf r}) 
G({\bf r} - {\bf r}') (q_{{\bf r}'} - f_{{\bf r}'})
\cr
  &+&\frac{J}{2L^{2}} \sum_{ \alpha=x,y } \sigma ^{2}_{ \alpha }
\end{eqnarray}
where
\begin{eqnarray}
\label{def2d}
\sigma_{x}&=&-2\pi \biggl[ L(q_{x1}-f_{x1})
+ \sum_{\bf r} (q_{\bf r}-f_{\bf r}) y \biggr]
\cr
\sigma_{y}&=&-2\pi \biggl[ L(q_{y1}-f_{y1}) 
- \sum_{\bf r} (q_{\bf r}-f_{\bf r}) x \biggr]
\cr
G({\bf r})&=&\frac{1}{N} \sum_{{\bf k} \neq 0}
 \frac{e^{i {\bf k} \cdot{\bf r}}-1}{4-2 \cos k_{x} -2 \cos k_{y}}
\end{eqnarray}
${\bf r}=(x,y)$ represents the coordinates of the sites of 
the dual lattice and $G({\bf r})$ is the lattice Green's function.
The topological charge, $q_{\bf r}$, is the circulation of the phase  
round the plaquette at ${\bf r}$,
$q_{\bf r} =\sum_{\Box {\bf r}} \phi_{ij}/2\pi$
and can be any integer subject to the neutrality condition 
$\sum_{\bf r} q_{\bf r}=0$. The frustration, $f_{\bf r}$, is
the circulation of $A_{ij}/2\pi$ around the plaquette at ${\bf r}$. The 
quantity $f_{x1}$ is the circulation of $A_{ij}/2\pi$  
round the whole torus on horizontal bonds of the plaquettes at $y=1$ and $q_{x1}$
is the coresponding circulation of the phase. $f_{y1}$ and $q_{y1}$ are defined similarly. 
Periodicity in the phases 
$\theta_{i}$ restricts $q_{x1}, q_{y1}$ to integers. 
\newline\indent
  In $3d$, the charge Hamiltonian becomes  
\begin{eqnarray}
\label{eq:h3d}
H=&2& \pi ^{2} J \sum_{{\bf r},{\bf r}'} ({\bf q}_{\bf r} - {\bf f}_{\bf r})
 \cdot ({\bf q}_{{\bf r}'} - {\bf f}_{{\bf r}'})G({\bf r} - {\bf r}')
\cr  
 &+& \frac{J}{2L} \sum_{\alpha=x,y,z} \sigma^{2}_{\alpha}
\end{eqnarray}
where
\begin{eqnarray}
\label{eq:def3d}
\sigma_{x} &=& \pi L^{-1} 
\sum_{\bf r} \Bigl\{ -z(q_{\bf r}^{y}-f_{\bf r}^{y}) 
                     +y(q_{\bf r}^{z}-f_{\bf r}^{z}) \Bigr\} + Q_{x}
\cr
Q_{x}&=&-\pi \sum_{\bf r} \Bigl(-zf_{\bf r}^{y}\delta_{y,1}
+yf_{\bf r}^{z}\delta_{z,1} \Bigr) +2 \pi L(q_{x1}-f_{x1}) 
\cr
G( {\bf r})&=&\frac{1}{N} \sum_{{\bf k} \neq 0} 
\frac{e^{i {\bf k} \cdot {\bf r} } -1}
{ 6 -2 \cos k_{x} -2 \cos k_{y} -2 \cos k_{z} }
\end{eqnarray}
with similar expressions for $\sigma_{y}, \sigma_{z}, Q_{y}, Q_{z}$ which are 
obtained by cyclic permutations of $(x,y,z)$.
Here, as in $2d$, $ {\bf r} = (x,y,z) $ are the coordinates of the dual lattice sites 
and $G({\bf r})$ is the Green's function. The 
charge ${\bf q}_{\bf r}$ and the frustration ${\bf f}_{\bf r}$
associated with the dual lattice at ${\bf r}$ 
become vector quantities with components  
$(q_{\bf r}^{x},q_{\bf r}^{y},q_{\bf r}^{z})$ and
$(f_{\bf r}^{x},f_{\bf r}^{y},f_{\bf r}^{z})$. Here $q_{\bf r}^{\alpha}$ 
and $f_{\bf r}^{\alpha}$ are the circulations of  
$\theta_{i}$ and $A_{ij}$ about the plaquette at ${\bf r}$ normal to the direction $\alpha$. 
 These vector charges  
satisfy $\nabla \cdot {\bf q_{r}} = 0 $ at each site 
${\bf r}$ and a neutrality condition as in $2d$.
\newline\indent  
 To find the true GS energy of the finite system of $L^{d}$ sites on a hypertorus,
we minimize the energy given by eq.(\ref{eq:h2d}) with respect to the
variables $q_{\bf r}$ and $f_{x1},f_{y1}$ in $2d$, and by eq.(\ref{eq:h3d}) in
$3d$.  
The lowest energy of the system is $2 \pi$ periodic in the twist
$\Delta_{\mu}\equiv 2\pi f_{\mu 1}$,   
with a minimum at some $\Delta_{\mu}^{0} $ 
which depends on the particular realization of disorder.
The lowest energy with $\Delta_{\mu} \neq \Delta_{\mu}^{0}$ includes
the excitation energy due to the twist $\Delta_{\mu} - \Delta_{\mu}^{0}$.
Adding a twist is equivalent to the gauge transformation 
$A_{ij} \rightarrow A_{ij} + \Delta_{\mu}/L$ on
all bonds in the direction $\mu$.
\newline\indent
We compute the domain wall energy $\Delta E(L)$ in two different ways. The 
first is with the usual periodic and antiperiodic BC, which correspond to twists ${\bf\Delta}
\neq{\bf\Delta^{0}}$ and ${\bf\Delta}+\pi{\bf\hat x}$ where ${\bf\Delta}$ is determined 
by the particular sample which we call a random twist (RT) measurement. The second is by
computing the true domain wall energy $\Delta E^{BT}(L)=E_{L}({\bf\Delta^{0}}+\pi{\bf\hat x})
-E_{L}({\bf\Delta^{0}})$ where $E_{L}({\bf\Delta^{0}})$ is obtained by minimizing the energy
with respect to the bulk charges $q^{\alpha}_{\bf r}$ and the twists $f_{1\alpha}$. The
domain wall is induced by $f^{0}_{1x}\rightarrow f^{0}_{1x}+1/2$ which corresponds to
$\Delta^{0}_{x}\rightarrow\Delta^{0}_{x}+\pi$ which we call a best twist (BT) measurement.
Note that, since $E_{L}({\bf\Delta^{0}})$ is the absolute energy minimum, $\Delta E^{BT}(L)\ge 0$
and the boundary contributions to the total energy proportional to $\sigma_{\alpha}^{2}\ge 0$
in eqs.(\ref{eq:h2d},\ref{eq:h3d}) must vanish when the best twist ${\bf\Delta^{0}}$ is applied.
These may be used as checks on the numerical algorithm. We use simulated annealing\cite{anneal1,anneal2} to
estimate the lowest energies, which is considerably more efficient than repeated quenches
to $T=0$. To minimize errors in the computation of domain wall energies, we used two random
number sequences to determine the annealing schedule and demanded that both sequences yield
the same GS energy, except for our largest system in $3d$ with $L=7$ because of time constraints.
The main source of error is statistical due to the averaging over disorder, but the ratio 
$\Delta E(L)/E_{L}({\bf\Delta^{0}})$ decreases rapidly with increasing $L$, especially in
$3d$ where this ratio is about $10^{-2}$ for $L=7$, which means that $E_{L}({\bf\Delta})$ must
be obtained almost exactly. This puts severe constraints on the accessible system sizes $L$.
\newline\indent
In $2d$, we compute the domain wall energy for sizes $2\le L\le 10$, taking averages over
about $10^{3}$ realizations of disorder for each $L$. The results of RT and BT measurements
are shown in fig.(1). The RT measurement gives $\theta^{RT}=-0.45\pm 0.015$ which agrees
with other groups, {\it{all}} of whom have used this measurement\cite{gg-simulation1,gg-simulation2,gg-simulation3,gg-simulation4,gg-simulation5,gg-simulation6}.
This agreement is not surprising as the only difference between these is in the computing power
and in the algorithms used. All are measuring the same quantity $\Delta E^{RT}(L)$.
The true defect energy obtained from the BT measurement gives $\theta^{BT}=-0.36\pm 0.010$
which is considerably larger. In $3d$, the system sizes are $2\le L\le 7$ with disorder 
averaging over $10^{3}$ realizations for $L\le 5$, $300$ for $L=6$ and $60$ for $L=7$. The
error in $\Delta E(L=7)$ is very large, but this point was included to check that it is
consistent with the behavior deduced from the smaller systems. The results are shown in
fig.(2). For the RT measurement, there seems to be a crossover around $L=5$ from a very
small value of $\theta^{RT}$ to a larger positive value as is also seen in 
\cite{gg-simulation6}, but our sizes do not allow any estimate of $\theta^{RT}$. On the
other hand, the BT measurement for these sizes is consistent with a power law scaling 
with a stiffness exponent $\theta^{BT}=+0.31\pm 0.010$ using sizes $L\le 6$, which is strong
evidence in favor of a superconducting glass phase at finite $T$ and of $d_{l}<3$. This is
also consistent with finite $T$ Monte Carlo results on the $3d$ gauge glass \cite
{gg-simulation1,gg-simulation2,gg-sim7} which indicate $T_{c}\sim O(J)$, which is difficult to
reconcile with the very small value of $\theta^{RT}$ which, if it were the stiffness exponent
for the $3d$ gauge glass, would imply $d_{l}\approx 3$ and a small value of $T_{c}/J$. If we
include the $L=7$ point, the best fit gives $\theta^{BT}=+0.30\pm 0.015$. Note that the errors
quoted here in $\theta^{BT,RT}$ come from a naive least squares fit to the data, and should not
be taken too seriously. The $L=7$ data is suspect because, in $10^{3}$ CPU hours on a Cray J90,
4 samples of a batch of 64 violated the BT condition $\Delta E^{BT}(L)\ge 0$ implying insufficient
annealing to reach the true energy minima. What data we have is entirely consistent with the
scaling form of eq.(\ref{eq:deltae}) with $\theta\approx +0.3$ with no sign of any deviation
from this. 
\newline\indent
We also studied the effects of screening on the domain wall energy using the BT measurement in 
$3d$. Screening of the interaction of charges is implemented by adding a term $\lambda^{-2}$,
where $\lambda$ is the screening length, to
the denominators of the Green's function of eq.(\ref{eq:h3d})\cite{screening}. The results are
also shown in fig.(2). We averaged over $10^{3}$ samples for $L=2,3,4$ and $250$ for $L=5$. For
the shorter screening lengths, screening is clearly a relevant perturbation and destroys
the ordered phase while for longer screening  lengths $\Delta E(L)$ seems to scale the same
way as the unscreened case but we expect there is a crossover to a negative stiffness exponent at
length scales $L$ which are inaccessible with our computing power. Our results are consistent
with those of Bokil and Young\cite{screening} who studied the screening question using the RT
measurement.
\newline\indent
The major result of this study is that one can, in principle, find the exact GS energy of a
{\it{random}} $XY$ system by a suitable choice of boundary conditions which are consistent with
the unknown GS. This is implemented for the gauge glass model in both $2d$ and $3d$ and it
is also argued that a domain wall is created from the GS by an appropriate change of the BC.
In the Coulomb gas representation of the system on a hypertorus, the BC are parametrized by $d$
numbers which are the circulations of $A_{ij}$ round loops enclosing the hypertorus. Once these
are known, a domain wall is induced by changing one of these by $\pi$. The size dependence of
the domain wall energy is computed and is found to scale rather accurately as $L^{\theta}$.
The values of the stiffness exponent $\theta$ are much larger than previous estimates and is
in accord with $d_{l}<3$. This also reconciles the finite $T$ Monte Carlo results with one's
physical intuition. We stress that the disagreement between the stiffness exponent $\theta^{BT}$
in this work and all previous estimates is because these measure $\theta^{RT}$ which is a quantity
whose meaning is unclear and is more likely to suffer from large corrections to scaling, especially
for the small $L$ values which can be simulated (see fig.(2) and ref.\cite{gg-simulation6}).  
We conjecture that they
would coincide if {\it{much}} larger values of $L$ could be reached. The stiffness exponent
$\theta^{BT}$ is larger than the $\theta^{ISG}\approx +0.2$ for the $3d$ Ising spin glass\cite{isg}
but we see no contradiction here as continuous variables can adjust to frustration more 
easily than Ising spins and one expects gauge glass order to resist distortions
better than Ising spin glass order.
   
  Computations were performed at the Theoretical Physics 
Computing Facility at Brown University. JMK thanks A. Vallat and B. Grossman for countless
conversations on gauge glasses when many of the ideas in this paper were proposed.

\begin{figure}
\center
\begin{minipage}{8.0cm}
\epsfxsize= 8.0cm \epsfbox{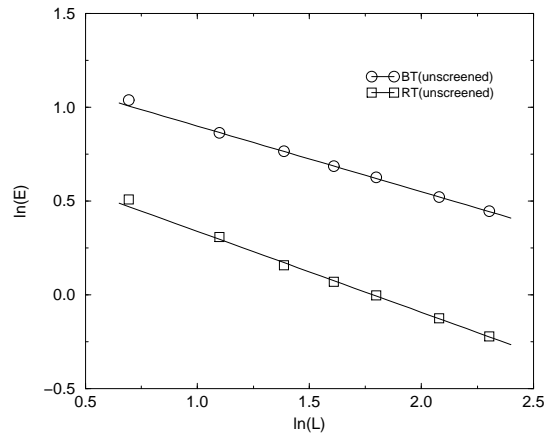}

\caption{Size $L$ dependence of domain wall energy in $2d$. Both RT and BT measurements are shown.
Solid lines are power law fits. Error bars are not shown if smaller than symbol size.}
\end{minipage}
\label{fig1}
\end{figure}

\begin{figure}
\center
\begin{minipage}{8.0cm}
\epsfxsize= 8.0cm \epsfbox{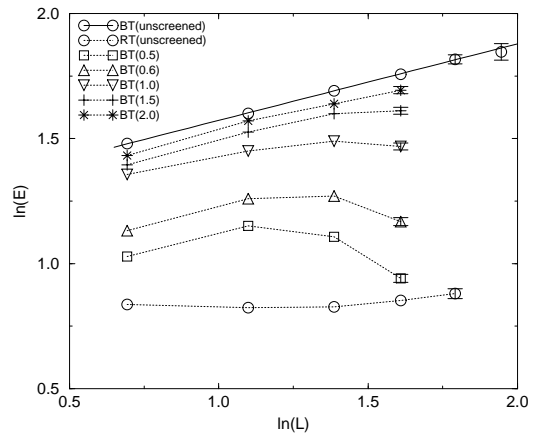}

\caption{$L$ dependence of domain wall energy in $3d$. Bottom curve is RT measurement for unscreened
interaction. All others are BT measurements. Topmost curve is unscreened case $L=2-7$. Other curves
are screened interactions with $\lambda$ decreasing from top to bottom. Solid lines are power law fits.
Dotted lines are guides for the eye.}
\end{minipage}
\label{fig2}
\end{figure}

\end{multicols}

\end{document}